\documentclass[fleqn,usenatbib]{mnras}

\usepackage{newtxtext,newtxmath}

\usepackage[T1]{fontenc}

\DeclareRobustCommand{\VAN}[3]{#2}
\let\VANthebibliography\thebibliography
\def\thebibliography{\DeclareRobustCommand{\VAN}[3]{##3}\VANthebibliography}

\usepackage{graphicx}	
\usepackage{amsmath}	

\usepackage{hyperref}

\title[Modeling AMR and mono-age MDF]{Modelling the broken age-[Fe/H] relation and metallicity distribution function of mono-age populations}

\author[Qinhao Shao et al.]{
Qinhao Shao,$^{1}$
    Jianhui Lian$^{1,2}$\thanks{E-mail: jianhui.lian@ynu.edu.cn (JL)}
\\
$^{1}$South-Western Institute for Astronomy Research, Yunnan University, Kunming, Yunnan 650091, People’s Republic of China \\
$^{2}${Yunnan Key Laboratory of Survey Science, Yunnan University, Kunming, Yunnan 650500, People’s Republic of China}
}

\date{Accepted XXX. Received YYY; in original form ZZZ}

\pubyear{\the\year{}}

\begin{document}
\label{firstpage}
\pagerange{\pageref{firstpage}--\pageref{lastpage}}
\maketitle

\begin{abstract}
This study investigates the chemical evolution history and radial migration in the Milky Way utilizing high-precision age measurements based on large-scale stellar spectroscopic surveys. Recent observational data have revealed a distinct break in the age–metallicity relation of stars in the solar neighborhood, suggesting that stellar populations of the Milky Way can be divided into two evolutionary sequences. Based on galactic chemical evolution models, we confirm that this break in the age–metallicity relation can be explained by a recent major gas accretion event, which diluted the metallicity in the disk and triggered a starburst, leading to the formation of an independent age–metallicity sequence. To assess the impact of stellar radial migration, we incorporate a simplified treatment of this process into our multi-zone chemical evolution model. Radial migration is modeled as a probabilistic diffusion process, characterized by a Gaussian distribution. By fitting the model‐predicted metallicity distribution functions (MDFs) of mono‐age stellar populations to observations, we infer a migration strength of $\sigma = 2.15^{+0.100}_{-0.103} \rm kpc$ for stars with ages between 2 and 6 Gyr located at 9.5-11.5 kpc, corresponding to an average radial displacement of 1.92 kpc over 10 Gyr. This result is lower than previous estimates of stellar radial migration, suggesting that the role of radial migration in shaping the chemical structure of the Milky Way is less important than previously estimated. 
\end{abstract}

\begin{keywords}
Galaxy: abundances -- Galaxy: evolution -- Galaxy: formation -- Galaxy: disc -- stars: abundances
\end{keywords}



\section{Introduction}\label{sec:intro}

The Milky Way (MW) serves as an exceptional chemical laboratory for studying galaxy formation and evolution. Understanding the evolutionary processes that have shaped the Milky Way into its present state is a fundamental question in astrophysics.

In their pioneering work, \citet{Chiappini_1997} proposed the influential “two-infall” model. This model posits that the Milky Way experienced two primary gas accretion phases: an initial rapid evolution phase leading to the formation of the thick disk, followed by a more prolonged evolution phase resulting in the thin disk. This seminal work sparked an ongoing debate: Was the Milky Way's chemical evolution governed by a single continuous accretion event or by two distinct episodes of gas infall? With the advent of large-scale sky surveys, numerous projects have provided an abundance of stellar spectroscopic data, such as the Large Sky Area Multi-Object Fiber Spectroscopic Telescope (LAMOST, \citeauthor{Zhao_2012} \citeyear{Zhao_2012}), Galactic Archeology with HERMES (GALAH, \citeauthor{slvai_2015} \citeyear{slvai_2015}), and the Apache Point Observatory Galactic Evolution Experiment (APOGEE, \citeauthor{Majewski_2017} \citeyear{Majewski_2017}). These high-quality spectroscopic datasets provide critical insights into the chemical compositions of stars, unveiling the distribution and temporal evolution of various elements across the Milky Way. At the same time, these surveys have introduced new challenges and opportunities, stimulating extensive discussions on the chemical evolution of the galaxy.

The bimodal abundance distribution of the Galactic disc has been well established by large spectroscopic surveys. In the $\mathrm{[\alpha/Fe]}$--$\mathrm{[Fe/H]}$ plane, stars are broadly separated into high- and low-$\alpha$ sequences, whose relative contributions vary with Galactocentric radius and height above the disc plane \citep[e.g.][]{Adibekyan_2012,Nidever_2014,Hayden_2015}. Several scenarios have been proposed to explain this chemical bimodality. Continuous or single-infall models can reproduce the two sequences through the combined effects of radial variations in chemical evolution and stellar radial migration \citep[e.g.][]{Schonrich_2009,Sharma_2021,Chen_2023}. Alternatively, two-infall models attribute the bimodality to distinct gas-accretion episodes, with the later accretion of metal-poor gas producing the low-$\alpha$ sequence \citep[e.g.][]{Chiappini_1997,Spitoni_2019,palla_2020,Lian_2020b}. Other studies have also explored explanations involving clumpy star formation or gas-rich merger events \citep[e.g.][]{Clarke_2019,Buck_2020}. More recently, \citet{Spitoni_2023} and \citet{palla_2024} investigated models involving an additional late gas-infall episode.

Due to the limited precision of stellar age determinations in earlier observational data, most previous studies have mainly focused on the $\mathrm{[\alpha/Fe]}$--$\mathrm{[Fe/H]}$ plane. Nevertheless, the age--metallicity relation (AMR) provides an equally important avenue for investigating the evolutionary history of the Milky Way. Classical chemical evolution models generally predict that stellar metallicity increases with time, implying that younger stars should, on average, be more metal-rich. However, observations have revealed substantial departures from such a simple monotonic relation. In particular, stars near solar metallicity tend to have younger mean ages than those at supersolar metallicities, producing a turnover or ``break'' in the AMR \citep{Anders_2017,Silva_2018,Feuillet_2019,Hasselquist_2019}. Similar features have also been identified with LAMOST data \citep{Xiang_2017,Wu_2018,Xiang_2022}, while \citet{Lian_2022} further showed that the broken age--[Fe/H] relation is present at $R>6$ kpc and becomes increasingly pronounced towards larger Galactocentric radii. This feature may arise either from the formation of younger, relatively metal-poor stars following dilution of the enriched interstellar medium or from radial migration that brings older, metal-rich stars from the inner disc to larger radii.

Recently, \citet{Xiang_2022} presented a catalog of $\sim$250,000 subgiant stars with high-precision age estimates. This sample revealed a broken AMR in the age–[Fe/H] plane, characterized by two distinct sequences. Such precise age constraints offer a powerful means of testing and refining Galactic chemical evolution models in the age–metallicity domain. Consequently, the first part of this work focuses on the age–[Fe/H] space, aiming to reproduce and interpret the observed broken AMR using a modified “two-infall” model following \citet{Lian_2020a, Lian_2020b}.

Stellar radial migration plays a significant role in shaping the chemical structure of the Milky Way \citep{Hayden_2015, Sharma_2021, Lian_2022, Lu_2024, Marques_2025, Bernaldez_2026, Bernaldez_2026b}. Radial changes in stellar orbits are generally associated with two processes: ``churning'' and ``blurring''. Churning refers to a change in a star's guiding radius through the exchange of angular momentum with non-axisymmetric structures, particularly spiral arms and the Galactic bar \citep{Sellwood_2002, Roskar_2008, Minchev_2010}. In contrast, blurring describes radial excursions around a nearly unchanged guiding radius caused by orbital heating \citep{Daniel_2019}. For the long-term redistribution of stellar populations across the Galactic disc, churning is expected to play a more important role than blurring, since it changes the stellar guiding radius through angular-momentum exchange, whereas blurring primarily increases the radial excursion around an approximately unchanged guiding radius \citep{Sellwood_2002,Schonrich_2009,Feltzing_2020}. Empirical modelling of the Milky Way disc further suggests that angular-momentum diffusion associated with churning is substantially stronger than radial heating associated with blurring \citep{Frankel_2020,Lian_2022}. Therefore, the radial migration considered in this work is primarily associated with changes in stellar guiding radii driven by churning. The availability of increasingly precise stellar ages, particularly from \citet{Xiang_2022}, now provides an opportunity to place tighter quantitative constraints on this process.

We adopt the method for implementing stellar radial migration proposed by \citet{Frankel_2018}, which approximates stellar radial migration as a Gaussian-like probability distribution. This approach simplifies the complexity of stellar migration by introducing a single free parameter, $\sigma$, which describes the expansion of the Gaussian distribution and represents the stellar radial migration strength. In this framework, stellar radial migration is treated as an instantaneous change in orbital radius \citep{Frankel_2018, Johnson_2021}. To constrain the radial migration strength, we analyze the metallicity distribution function (MDF) of mono-age stellar populations at different galactic locations \citep{Lian_2022}. In the absence of stellar radial migration, the MDF of mono-age populations at a given location should remain narrow and well-defined. However, radial migration broadens and skews the MDF, mixing stars from different birth radii. By comparing the observed MDF of mono-age stellar populations with the MDF predicted by our model under various radial migration strengths, we can place constraints on the degree of radial migration in the Milky Way.

The remainder of this paper is organized as follows: In §2, we introduce the observational sample used in this study. In §3, we describe the construction of our chemical evolution model, including model parameters, implementation of stellar radial migration, and nucleosynthesis prescriptions. In §4, we compare our model predictions with observational data and derive constraints on the radial migration strength. In §5, we discuss our results and the limitations of our approach. Finally, in §6, we summarize our main conclusions.

\section{Sample} \label{sec:data}
Gaia Early Data Release 3 (eDR3; \citealt{Gaia_2021}), published by the European Space Agency's Gaia mission, offers precise astrometric measurements, including stellar positions, parallaxes, and proper motions, covering extensive regions of the Milky Way's stellar populations. The Gaia satellite achieves this through repeated sky scans using a high-precision optical system specifically designed for astrometry and photometry.The Large Sky Area Multi-Object Fiber Spectroscopic Telescope (LAMOST; \citealt{Cui_2012,Zhao_2012}) is a ground-based spectroscopic survey facility at the Xinglong Observatory. Equipped with a multi-object fiber spectroscopic system, LAMOST can simultaneously observe up to 4,000 celestial objects, enabling large-scale spectroscopic studies of the Milky Way's stellar populations. LAMOST Data Release 7 (DR7) provides extensive spectral data, including stellar parameters such as radial velocities, effective temperatures, surface gravities, and metallicities.

In this study, we utilize the sample from \citet{Xiang_2022} to constrain our chemical evolution model. Their data set was constructed by combining Gaia eDR3 astrometric data with LAMOST DR7 spectroscopic observations. Xiang et al. identified a sample of approximately 250,000 subgiant stars spanning vast regions of the Milky Way. The age range of this sample is $1.5 \text{Gyr} < \tau < 13.8 \text{Gyr}$, and the metallicity range is $-2.5 < [\text{Fe}/\text{H}] < 0.4$. Additionally, they derived high-precision stellar ages using Bayesian fitting with Yonsei–Yale (YY) stellar isochrones, with estimated uncertainties around $7.5\%$. The exceptional precision of these age estimates provides stringent constraints on the age-metallicity relation (AMR) in our model. The left subplot of Fig. \ref{fig:amr} shows the distribution of this sample in the age-metallicity diagram.

Based on this sample, we can further investigate the relationship between stellar age and metallicity within the Milky Way and use this relationship to constrain chemical evolution models. The chemical composition measured in subgiant star spectra directly reflects the chemical composition of the interstellar medium at the time of their formation. Therefore, by statistically analyzing the metallicity distribution of stars of different ages, we can reconstruct the chemical enrichment processes of the Milky Way at various epochs. As shown in the left panel of Fig. \ref{fig:amr}, the age-metallicity relation of the Milky Way is characterized by a break. This information allows us to delve into the chemical evolution history of the Milky Way, revealing whether it has evolved into its current complex chemical structure through two independent accretion events. Moreover, the high-precision age information also supports the study of the metallicity distribution function of mono-age stellar populations. Analyzing the metallicity distribution function of mono-age stellar populations can help constrain the stellar radial migration strength, thereby assessing the role of stellar radial migration in the formation of the Milky Way's current chemical structure.

\section{The Chemical Evolution Model} \label{sec:model}
The chemical evolution model employed in this study is designed to trace the chemical evolution of galaxies on both global and local scales. This model successfully reproduced key chemical properties of the Milky Way, including the metallicity trends in its outer disk, the bimodal $\mathrm{[\alpha/Fe] - [Fe/H]}$ distribution in the inner disk and bulge of the Galaxy \citep{Lian_2020a,Lian_2020b}. In this work, we adopt a multi-zone chemical evolution model extended from the original one-zone framework, with the effects of stellar radial migration incorporated. An initial implementation of this framework was presented in \citet{Lian_2026}. The following sections outline the key components of our model.

\subsection{Parameters} \label{subsec:parameter}
Here, we briefly introduce the basic assumptions and input parameters of our model. The Galactic chemical evolution model presented in this work is an updated version of the framework described in \citet{Lian_2020a,Lian_2020b}. The model follows the chemical evolution within a region of $1~{\rm kpc}^{2}$, which is adopted as the normalization area. We combine 12 such chemical evolution zones to model the radial evolution of the Galactic disc over $0<R<12.5$~kpc. Because the four innermost zones at $R<4.5$~kpc adopt identical parameters, they are combined into a single row in Table~\ref{tab:pa}, leaving nine distinct parameter sets for the 12 radial zones.

We prescribe the gas-accretion and star-formation histories independently for each radial region. The gas-accretion history is divided into four evolutionary phases. During the initial phase ($t \leq t_{\rm ini}=2.5$~Gyr), the gas accretion rate is kept constant at $A_{\rm initial}$. During the subsequent secular evolution phase ($2.5<t\leq6$~Gyr), the gas accretion rate declines exponentially from its value at the end of the initial phase,

\begin{equation}
A(t)=A(t_{\rm ini})
\exp\left[-\frac{t-t_{\rm ini}}{\tau_{\rm secular}}\right],
\end{equation}

where the e-folding timescale is fixed at $\tau_{\rm secular}=0.5$~Gyr. As the external gas supply rapidly declines during this phase, the gas reservoir generally decreases as gas continues to be consumed by star formation.

At $t=6$~Gyr, a second gas-accretion episode is introduced in most radial regions. During this phase ($6<t\leq7$~Gyr), the gas accretion rate is kept constant at $A_{\rm second}$. After $t=7$~Gyr, the external gas accretion is terminated. In the innermost regions, where $A_{\rm second}=0$, only the initial gas-accretion episode is included.

The evolution of the gas reservoir is calculated self-consistently from the balance among external gas accretion, gas consumption by star formation, and gas returned to the interstellar medium through stellar evolution. Therefore, the gas surface density, $\Sigma_{\rm gas}$, does not simply follow the prescribed gas-accretion history, but instead reflects the combined effects of these processes.

We adopt the Kennicutt--Schmidt (KS) star-formation law \citep{kennicutt_1998} to describe star formation in each radial region. The star formation rate surface density is calculated as

\begin{equation}
\Sigma_{\rm SFR} =
C_{\rm SFL}\times 2.5\times10^{-4}
\left(0.75\,\Sigma_{\rm gas}\right)^{n_{\rm KS}},
\end{equation}

where $\Sigma_{\rm gas}$ is the gas surface density and $n_{\rm KS}=1.4$. The dimensionless coefficient $C_{\rm SFL}$ controls the star formation efficiency at a given gas surface density, such that a larger $C_{\rm SFL}$ corresponds to a higher star formation rate for the same gas reservoir.

The evolution of $C_{\rm SFL}$ is likewise divided into four phases. During the initial phase, $C_{\rm SFL}$ is fixed at $C_{\rm SFL,initial}$. During the secular evolution phase, $C_{\rm SFL}$ declines exponentially from $C_{\rm SFL,initial}$ with an e-folding timescale of $\tau_{\rm SFE}=1$~Gyr, until it reaches the prescribed lower limit $C_{\rm SFL,secular}$,

\begin{equation}
C_{\rm SFL}(t)=
\max\left[
C_{\rm SFL,initial}
\exp\left(-\frac{t-t_{\rm ini}}{\tau_{\rm SFE}}\right),
C_{\rm SFL,secular}
\right].
\end{equation}

During the second gas-accretion phase, $C_{\rm SFL}$ is set to $C_{\rm SFL,second}$, while after the second accretion episode it is fixed at $C_{\rm SFL,post}$.

The adopted values of $A_{\rm initial}$, $A_{\rm second}$, $C_{\rm SFL,initial}$, $C_{\rm SFL,secular}$, $C_{\rm SFL,second}$, and $C_{\rm SFL,post}$ vary with Galactocentric radius and are summarized in Table~\ref{tab:pa}.

\begin{table*}
\centering
\footnotesize
\caption{Model parameters adopted in different Galactocentric radial ranges.}
\label{tab:pa}
\begin{tabular}{ccccccc}
    \hline
    \hline
    $R$ & $A_{\rm initial}^{a}$ & $A_{\rm second}$ &
    $C_{\rm SFL,initial}$ & $C_{\rm SFL,secular}$ &
    $C_{\rm SFL,second}$ & $C_{\rm SFL,post}$ \\
    (kpc) &
    ($M_{\odot}\,{\rm yr}^{-1}\,{\rm kpc}^{-2}$) &
    ($M_{\odot}\,{\rm yr}^{-1}\,{\rm kpc}^{-2}$) & & & & \\
    \hline
0--4.5     & 0.1    & 0     & 1 & 0.3 & 0.3  & 0.3  \\
4.5--5.5   & 0.04   & 0.015 & 1 & 0.4 & 0.4  & 0.3  \\
5.5--6.5   & 0.02   & 0.025 & 1 & 0.4 & 0.4  & 0.3  \\
6.5--7.5   & 0.02   & 0.04  & 2 & 0.5 & 0.5  & 0.3  \\
7.5--8.5   & 0.007  & 0.025 & 3 & 0.5 & 0.5  & 0.3  \\
8.5--9.5   & 0.005  & 0.03  & 4 & 0.5 & 0.5  & 0.25 \\
9.5--10.5  & 0.003  & 0.032 & 5 & 0.5 & 0.5  & 0.2  \\
10.5--11.5 & 0.0015 & 0.035 & 6 & 0.3 & 0.45 & 0.17 \\
11.5--12.5 & 0.0008 & 0.04  & 8 & 0.3 & 0.45 & 0.1  \\
    \hline
\multicolumn{7}{l}{$^{a}$ Gas accretion rate surface density.}
\end{tabular}
\end{table*}

Our model adopts the stellar initial mass function (IMF) of \citet{Kroupa_2001}. Given that the efficiency of metal outflows is inversely correlated with galaxy mass \citep{Chisholm_2018,Lian_2018,lin_2023}, and considering the relatively high stellar mass of the Milky Way, we assume that significant galactic outflows are absent in the present model.

\subsection{Stellar radial migration}\label{subsec:migration}
Theoretical models and observational studies have increasingly highlighted the crucial role of stellar radial migration in shaping the chemical structure of the Milky Way (e.g., \citet{Johnson_2021}). To account for this effect, we incorporate stellar radial migration into our model, allowing for interactions and mixing among the 12 radial regions. This approach allows us to more accurately reproduce the radial chemical trends observed in the Milky Way.

Our implementation of stellar radial migration follows the diffusion-based framework proposed by \citet{Frankel_2018}. According to this model, radial migration occurs as a result of stars gaining or losing angular momentum through interactions with the bar or spiral arms, thereby leading to their radial migration. Assuming a star is born at radius $R_0$ and undergoes migration over a time period $t$, the probability of finding it at radius $R$ is given by the following equation:

\begin{equation}
P(R|R_0, t) = C \exp \left( -\frac{(R - R_0 + \sigma^2 t/2 R_d t_0)^2}{2\sigma^2 t / t_0 } \right),
\label{eqn:migration}
\end{equation}

where \(R_d\) represents the scale length of each mono-age stellar population, for which we adopt the best-fitting results from \citet{zhang_2026}. The term \(\sigma^2 t/(2R_d t_0)\) represents a drift term to the otherwise symmetric Gaussian diffusion. A purely symmetric redistribution in radius can produce a net outward transfer of stars in an exponential disc, because the number of stars available to migrate outward from the inner disc is larger than that available to migrate inward from the outer disc. \citet{Frankel_2018} noted that such a simple Gaussian parameterization can therefore result in a presumably unphysical net outward motion. The drift term shifts the centre of the redistribution kernel slightly inward and compensates for this global bias; in the corresponding angular-momentum diffusion formulation, such a term is introduced to approximately conserve the total angular momentum of an exponential disc (\citet{Sanders_2015,zhang_2026}). The parameter \(\sigma\) specifies the width of the Gaussian redistribution function at the reference timescale \(t_0=8\) Gyr and therefore characterizes the strength of stellar radial migration. Here, \(t\) denotes the elapsed time since stellar birth, i.e. the stellar age. The constant $C$ is a normalization factor, given by the equation:

\begin{equation}
C^{-1} = \sigma \sqrt{ \frac{\pi}{2} \frac{t}{t_0}} \left( \text{erf} \left( \frac{R_0 - \sigma^2 t/2 R_d t_0}{\sigma \sqrt{2} \sqrt{t / t_0}} \right) + 1 \right)
\label{eqn:C}
\end{equation}

Based on the method of implementing stellar radial migration. In our best-fit model, the radial migration strength $\sigma = 2.15$ kpc.

\subsection{Nucleosynthesis Prescription}\label{subsec:nucle}
A model of Galactic chemical evolution aims to describe the production, distribution, and temporal evolution of elements within the Galaxy, in which the nucleosynthesis prescriptions (yields) from various stellar evolutionary stages play a decisive role. The metallicity of the Milky Way is the result of the accumulation of elements synthesized by multiple generations of stars through internal nuclear reactions, which are then released through processes, including stellar winds and supernova explosions. A crucial input parameter in the model is the nucleosynthesis yield of stars within each mass range, as it directly governs the spatiotemporal distribution and evolutionary trends of different chemical elements.

Our model incorporates enrichment from asymptotic giant branch (AGB) stars, Type Ia supernovae (SN-Ia) and Type II supernovae (SN-II). We adopted the metallicity-dependent AGB yields from \citet{Cristallo_2015}, SN-Ia yields from \citet{Iwamoto_1999} and SN-II yields from \citet{Limongi_2018}. For SNe Ia, we implement a power-law delay time distribution (DTD) following \citet{Maoz_2012}, which takes the form:

\begin{equation}
DTD = a(\tau / \text{Gyr})^{-1}
\label{eqn:DTD}
\end{equation}

where $\tau$ represents the delay time since the formation of the progenitor binary system. The normalization constant $a$ is determined by the following equation:

\begin{equation}
\textstyle \int_{\tau_{min}}^{\tau_{max}} DTD(\tau) d\tau = 1
\label{eqn:a}
\end{equation}

where $\tau_{\min}=0.15$ Gyr and $\tau_{\max}=21$ Gyr denote the minimum and maximum delay times, respectively. We adopt $\tau_{\min}$ following the fiducial value used by \citet{Andrews_2017}.

At each evolutionary timestep, the model calculates the mass of gas and individual chemical elements returned to the interstellar medium (ISM) through different stellar enrichment channels, including AGB stars, SNe Ia, and SNe II, according to the adopted stellar yields and enrichment timescales described above. These ejecta are combined with the existing gas reservoir to update the elemental abundances of the ISM, from which the corresponding abundance ratios are calculated. Newly formed stars are assumed to inherit the chemical composition of the ISM at the time of their formation.

\section{Results} \label{sec:result}
In this section, we present the predictions of our model in detail and compare them with the observational sample.

\subsection{Age-[Fe/H] relation}\label{subsec:broken}
In generating mock catalogs from the model, we incorporated an age uncertainty of 0.03 dex and a metallicity uncertainty of 0.07 dex (consistent with the observational sample). To determine the number of stars sampled per evolutionary radial bin, we adopted the radial variation of the Galactic surface mass density profile from \citet{Lian_2025}.

\begin{figure*}
\centering
\includegraphics[width=16cm]{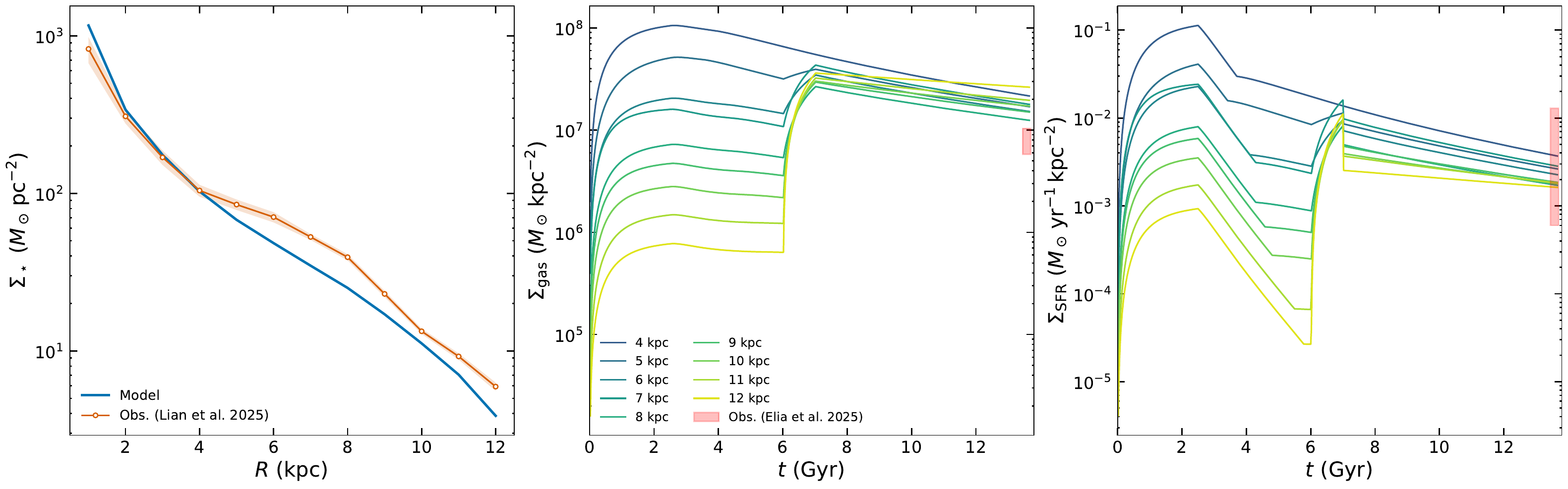}
\caption{Radial and temporal evolution of the fiducial two-infall model. Left: present-day stellar surface mass-density profile as a function of Galactocentric radius. The blue line shows the model prediction, while the orange points and shaded region indicate the observational constraint from \citet{Lian_2025}. Middle: evolution of the gas surface density, \(\Sigma_{\rm gas}\), at different Galactocentric radii. Right: corresponding evolution of the star formation rate surface density, \(\Sigma_{\rm SFR}\). The red shaded regions indicate the adopted present-day observational constraints, from \citet{Elia_2025}. In the middle and right panels, \(t\) denotes the time elapsed since the onset of Galactic evolution. The rapid increase in \(\Sigma_{\rm gas}\) and \(\Sigma_{\rm SFR}\) around \(t\simeq6\) Gyr is associated with the second gas-accretion episode.}
\label{fig:info}
\end{figure*}

We first examine the basic radial and temporal properties of our fiducial two-infall model. Fig.~\ref{fig:info} shows the present-day stellar surface mass-density profile, together with the evolution of the gas and star formation rate surface densities at different Galactocentric radii. The predicted stellar surface mass density decreases systematically with increasing Galactocentric radius and broadly follows the observed radial profile. During the early evolutionary phase, the gas reservoir builds up rapidly and subsequently declines as gas is consumed by star formation. At a Galactic evolutionary time of \(t \simeq 6\) Gyr, the onset of the second gas accretion episode leads to a rapid increase in \(\Sigma_{\rm gas}\) across most of the Galactic disc. Through the adopted Kennicutt–Schmidt relation, this renewed gas supply also produces a corresponding enhancement in \(\Sigma_{\rm SFR}\). The inner disc does not exhibit this feature because only the initial gas-accretion episode is adopted in this region. However, the model overpredicts the present-day gas surface density compared with the observational constraint, indicating a limitation of the current model.

\begin{figure*}
\centering
\includegraphics[width=16cm]{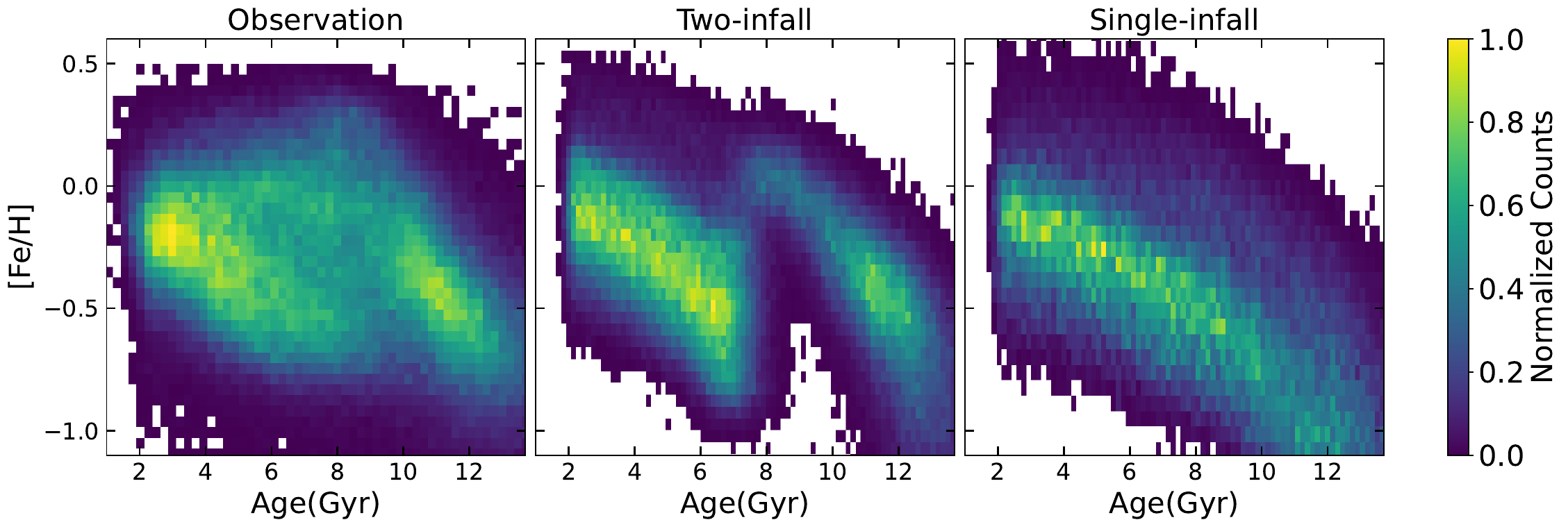}
\caption{Comparison of the distributions in the age--metallicity plane for the observations (left panel), the two-infall model (middle panel), and the single-infall model (right panel). Both model predictions include stellar radial migration with a migration-strength parameter of $\sigma=2.2$ kpc, defined at the reference timescale $t_0=8$ Gyr.}
\label{fig:amr}
\end{figure*}

To assess whether the simulated stellar populations can statistically reproduce the observed age–metallicity relation, we resampled the simulated catalogues within individual radial bins. To ensure a more consistent comparison with the observations, we excluded stars younger than 2 Gyr, which constitute only a negligible fraction of the observational sample. In each radial bin, we randomly selected the same number of stars as in the corresponding radial bin of the observed sample. The resulting mock subsamples were subsequently combined to construct a global Galactic sample, allowing a direct statistical comparison between the observations and the model predictions. For comparison, we also constructed a single-infall model incorporating stellar radial migration and applied exactly the same sampling procedure as for the two-infall model.

Fig.~\ref{fig:amr} compares the distributions in the age–metallicity plane for the observational data, the two-infall model, and the single-infall model, shown in the left, middle, and right panels, respectively. The two-infall model successfully reproduces the broken age–metallicity relation observed in the data, with the transition occurring at approximately 8 Gyr ago. The observed distribution shows two distinct sequences in the age–metallicity diagram, both of which are well recovered by the two-infall model. By contrast, the single-infall model fails to reproduce this feature and instead predicts only a single, continuous evolutionary track.

Overall, the two-infall model is broadly consistent with the observed age–metallicity distribution. However, the observed young sequence exhibits a pronounced density enhancement at ages of 2--4~Gyr, for which no corresponding overdensity is present in the model prediction. In addition, the relative density of the old sequence predicted by the model is lower than that observed. These discrepancies may be partially related to selection effects in the observational data. Another limitation is that the two-infall model does not predict stars with ages of approximately 4--8 Gyr and metallicities between about $-0.1$ and $0.4$ dex. Therefore, the current two-infall model cannot fully explain the origin of this stellar population. \citet{zhang_2025} suggested that these stars may have migrated from the inner Galaxy as the corotation resonance radius of the Galactic bar expanded outward.

\begin{figure*}
\centering
\includegraphics[width=16cm]{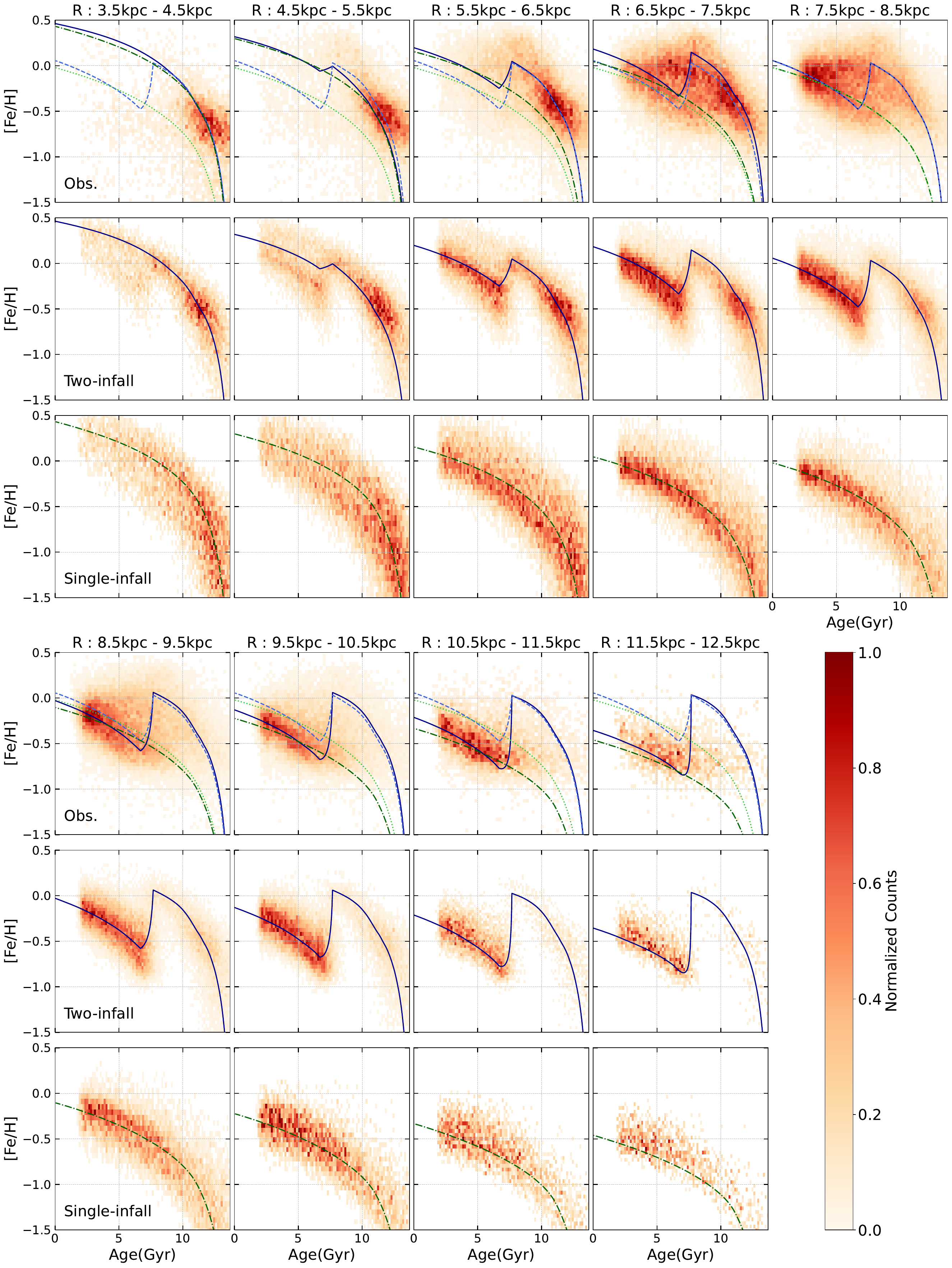}
\caption{Comparison of the observed distributions (first row), the distributions predicted by the two-infall model (second row), and those predicted by the single-infall model (third row) in the age--[Fe/H] plane at different Galactocentric radii. Both model predictions include stellar radial migration with a migration-strength parameter of $\sigma=2.2$ kpc, defined at the reference timescale $t_0=8$ Gyr. Dark blue solid and light blue dashed lines indicate the local and solar-neighbourhood evolutionary tracks predicted by the two-infall model, respectively. Dark green dotted and light green dash-dotted lines correspond to the local and solar-neighbourhood evolutionary tracks predicted by the single-infall model, respectively.}
\label{fig:amr_dr}
\end{figure*}

To further assess the validity of the model, we divided the observational sample into the same radial bins as adopted in the simulations and compared the corresponding age–metallicity distributions, as shown in Fig.~\ref{fig:amr_dr}. The first row presents the observational data in different radial bins, the second row shows the predictions of the two-infall model, and the third row displays the results of the single-infall model. Different columns correspond to different radial regions, as indicated in the figure. Since the model parameters are identical over the radial range of $0<R<4.5$ kpc, only the results for the $3.5<R<4.5$ kpc bin are shown.

The dark blue solid line represents the local evolutionary track predicted by the two-infall model, while the light blue dashed line denotes the corresponding evolutionary track in the solar neighbourhood. Similarly, the dark green dotted line indicates the local evolutionary track predicted by the single-infall model, and the light green dash-dotted line represents its prediction for the solar neighbourhood. As shown in the figure, the two-infall model successfully reproduces the observed break in the age–metallicity relation at an age of approximately 8 Gyr, and captures the bimodal sequence distribution seen near the solar vicinity. By contrast, the single-infall model provides a reasonable description of the observed age–metallicity distribution mainly in the inner and outer radial regions, but fails to reproduce the bimodal sequences observed around the solar neighbourhood.

Nevertheless, the two-infall model does not account for the origin of stars with ages of approximately 3--8 Gyr and metallicities around 0 dex in the radial range of $5.5<R<10.5$ kpc, particularly near the solar vicinity. As discussed above, these stars may have originated in the inner Galaxy and subsequently migrated to the solar neighbourhood as a result of the outward shift of the corotation resonance radius \citep{zhang_2025}. In addition, the two-infall model does not accurately reproduce the number-density distribution along the young sequence.

\subsection{MDF of mono-age populations}\label{subsec:MDF}
We constrain stellar radial migration by comparing the metallicity distribution functions (MDFs) of mono-age populations between the observed and model-predicted samples. The underlying rationale is that, in the absence of radial migration and intrinsic abundance scatter, stars of the same age formed within a given radial region should have nearly identical metallicities, and their MDF would therefore approach a delta function. However, stellar radial migration is expected to operate in the Galactic disc, where a large-scale radial metallicity gradient is present. As a consequence, mono-age populations observed at a given radius may include stars that migrated from different birth radii, where the interstellar medium had different metallicities at the same epoch. This process broadens the MDF and may also introduce asymmetry or skewness. The width and shape of the MDF can therefore be used to constrain the strength of stellar radial migration. A similar approach was adopted by \citet{Lian_2022}.

For this analysis, we focus on stars in the radial ranges of $9.5<R<10.5$ kpc and $10.5<R<11.5$ kpc, and divide them into four age bins: 2--3 Gyr, 3--4 Gyr, 4--5 Gyr, and 5--6 Gyr. These radial ranges were selected for several reasons. First, regions at $R<9.5$ kpc are dominated by the older population, while the solar vicinity is affected by an intermediate-age, metal-rich population whose evolutionary history remains unclear. Second, the observational sample becomes relatively sparse beyond $R=11.5$ kpc. The selected age range is also motivated by the model predictions and the observational sample size. Stars older than 6 Gyr are mainly associated with the older population, and the model predicts a metallicity dilution event at an age of approximately 6.7 Gyr, which would complicate the inference of radial migration strength. Stars younger than 2 Gyr are excluded because the number of observed stars in this age range is limited.

\begin{figure*}
\centering
\includegraphics[width=16cm]{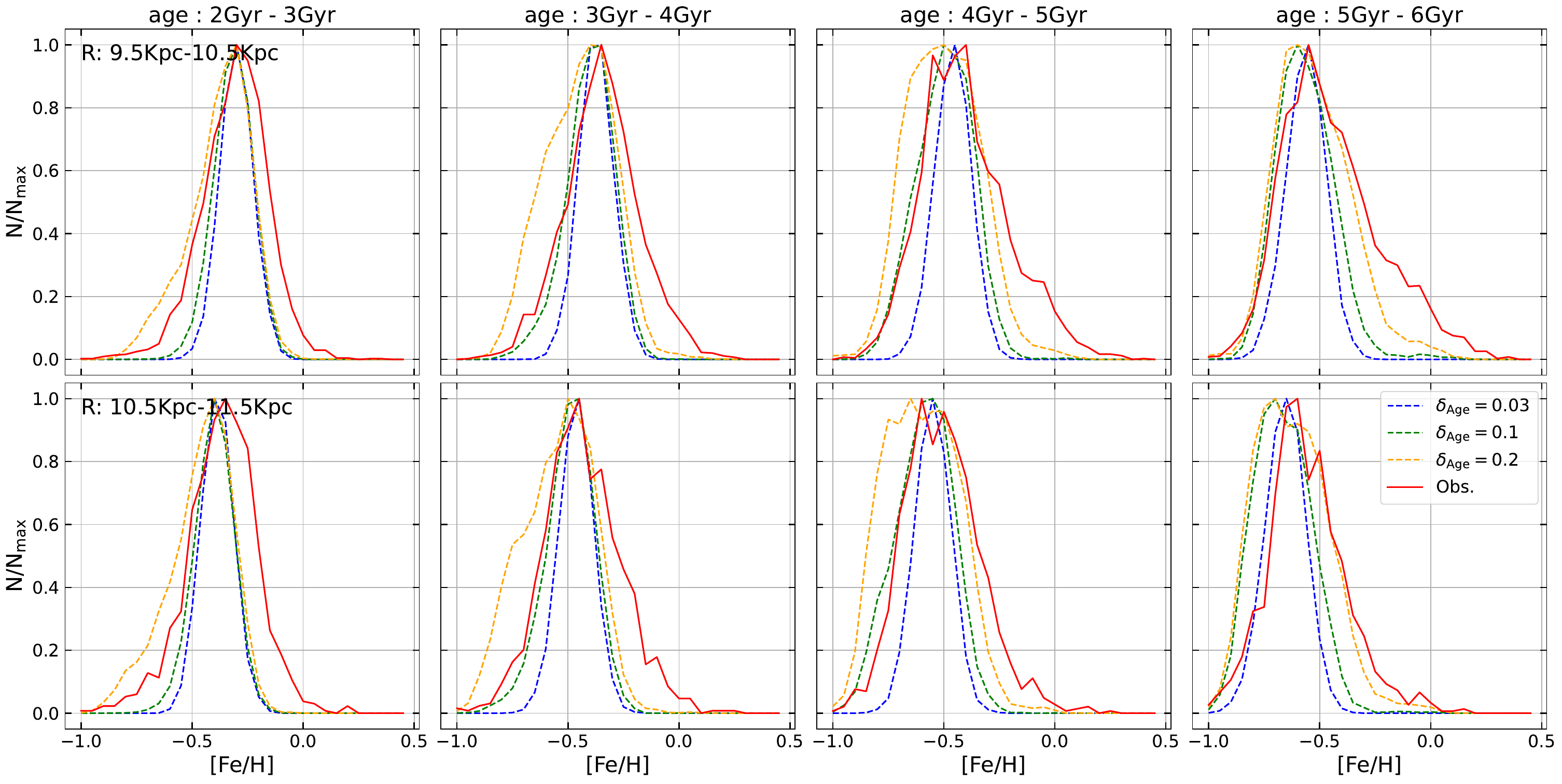}
\caption{Comparison of MDFs without considering stellar radial migration, by taking into account different age errors, with the observational sample. The first row shows the radial bin of 9.5 kpc–10.5 kpc, the second row 10.5 kpc–11.5 kpc. The red solid line represents the MDF of the observational sample, while the blue, green, and orange dashed lines correspond to the MDFs of model-predicted samples with age errors of 0.03, 0.1 and 0.2.}
\label{fig:delta}
\end{figure*}

We first examined the MDFs predicted by the model without stellar radial migration and compared them with the observation sample. Figure \ref{fig:delta} presents these comparisons, where the solid red line represents the observed MDFs, and the dashed lines correspond to model predictions with different age errors (0.03, 0.1, and 0.2 dex, shown in blue, green, and orange, respectively). We use this figure to examine whether the broadening observed in the MDF can be explained by increased age uncertainties. As shown in Fig. \ref{fig:delta}, introducing age errors into the model-predicted sample fails to yield a satisfactory reproduction of the observed MDF. Therefore, merely incorporating age uncertainties cannot substantially improve the agreement between the model and observations, indicating that age uncertainty affects the MDF distribution but is not the dominant cause of the discrepancy.

\begin{figure*}
\centering
\includegraphics[width=16cm]{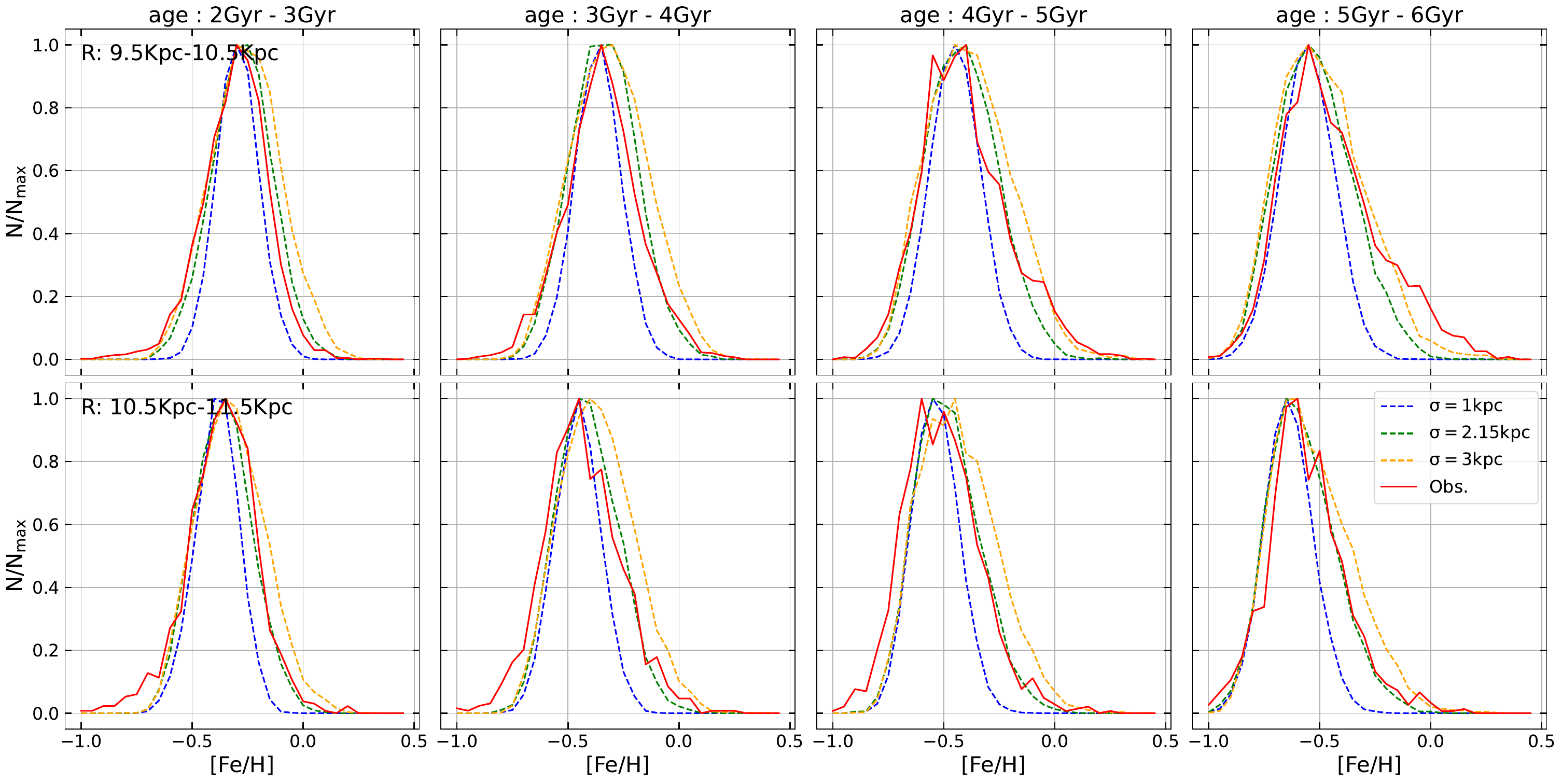}
\caption{Comparison of MDFs of model-predicted samples with varying stellar radial migration strengths and the observational sample. The first row shows the radial bin of 9.5 kpc–10.5 kpc, the second row 10.5 kpc–11.5 kpc. 
The red solid line represents the MDF of the observational sample, while the blue, green and orange dashed lines represent model-predicted MDFs with stellar radial migration strengths of 1 kpc, 2.15 kpc and 3 kpc.}
\label{fig:sigma}
\end{figure*}

Next, we explored the impact of varying stellar radial migration strengths on MDFs. Fig. \ref{fig:sigma} shows a comparison between the MDFs of the model-predicted MDFs with different stellar radial migration strengths and the MDFs of the observational sample. The red solid line represents the MDF of the observational samples. We set three different stellar radial migration strengths, represented by the blue dashed line for a stellar radial migration strength of 1 kpc, the green dashed line for 2.15 kpc and the orange dashed line for 3 kpc.

As illustrated in Fig. \ref{fig:sigma}, the model can better reproduce the observed MDF when stellar radial migration is included. However, for stars at 9.5–10.5 kpc with ages of 4–6 Gyr, the model underestimates the metal-rich end of the MDF. This discrepancy may arise from the simplified treatment of stellar radial migration in our chemical evolution model. Although the adopted diffusion kernel includes a drift term to account for the global outward bias associated with the decreasing stellar density profile, it does not explicitly describe coherent and strongly asymmetric migration driven by individual non-axisymmetric structures. In particular, \citet{zhang_2026} showed that the evolving corotation resonance of the Galactic bar can preferentially transport stars outward from the inner Galaxy.

To quantify the goodness of fit, we calculate the $\chi^2$ differences between the MDFs of model-predicted and observational sample, as defined by Eq.\ref{eqn:chi_2}.

\begin{equation}
\chi^2 = (MDF_{predicted} - MDF_{observed})^2
\label{eqn:chi_2}
\end{equation}

\begin{figure}
\centering
\includegraphics[width=\columnwidth]{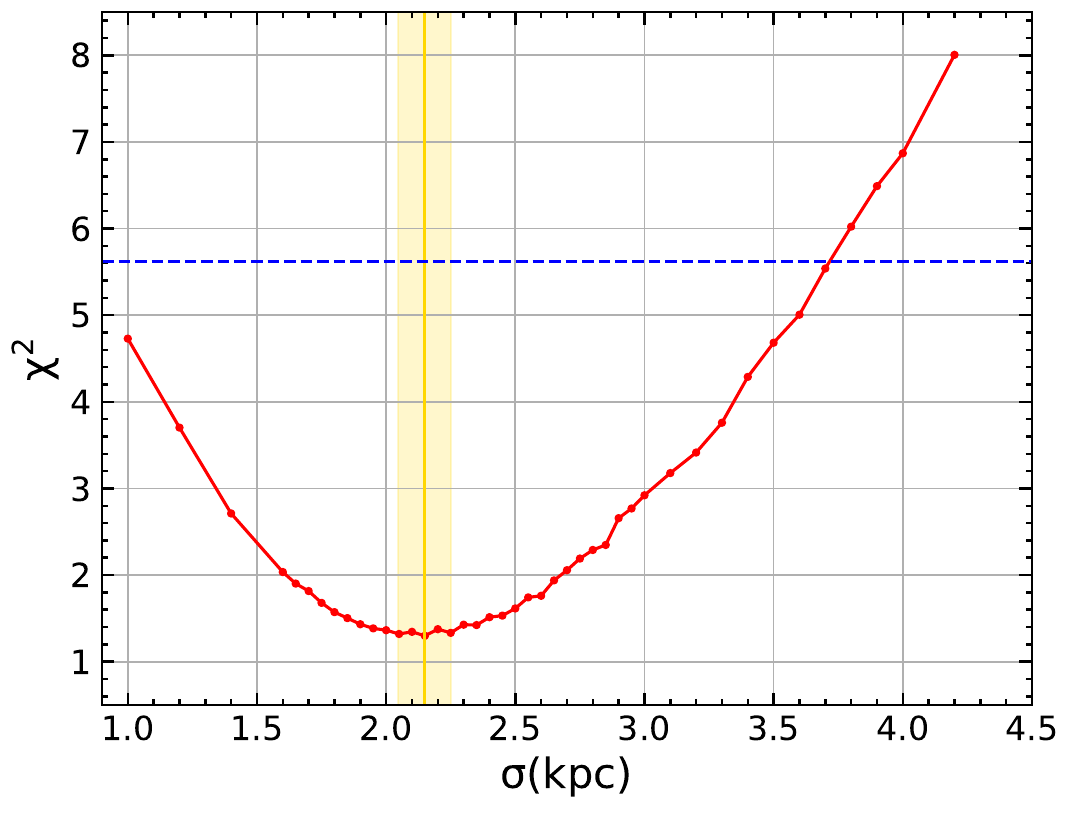}
\caption{The differences between model-predicted and observed MDFs for various migration strengths. The orange solid line represents the optimal stellar radial migration strength, while the orange rectangular region represents its uncertainty range. The blue dashed line represents the minimum $\chi^2$ value obtained for the model without stellar radial migration but with different age errors.}
\label{fig:chi_2}
\end{figure}

Fig. \ref{fig:chi_2} shows the obtained $\chi^2$ values for varying stellar radial migration strengths. The observational data impose strong constraints on lower migration strengths, with relatively weaker constraints on higher strengths. This method yields an optimal stellar radial migration strength of 2.15 kpc (represented by the orange solid line in the figure), corresponding to the lowest $\chi^2$ value of 1.30. This optimal stellar radial migration strength corresponds to an average stellar migration of 1.92 kpc over 10 Gyr. We also calculated the $\chi^2$ values under different age errors without stellar radial migration using the same method, with the minimum $\chi^2$ value being 5.63 (represented by the blue dashed line in the figure), which is significantly higher than the chi-square value corresponding to the optimal stellar radial migration model. Using Monte Carlo simulations and assuming Poisson errors for the observed MDF, we adopted a 
1 $\sigma$ error and derived an uncertainty of +0.100 kpc and -0.103 kpc for the best‑fit stellar radial migration strength (indicated by the orange shaded region in the figure).

\section{Discussion} \label{sec:discussion}

\subsection{Origin of the broken age-metallicity relation}\label{subsec:AMR}
The high-precision stellar ages provided by \citet{Xiang_2022} reveal a pronounced break in the age-metallicity relation (AMR), separating the Galactic disc into an older and a younger sequence. Our two-infall model provides a natural explanation for this structure. Following the early phase of chemical enrichment, a later accretion episode introduces relatively metal-poor gas into the disc, diluting the interstellar medium (ISM) and subsequently triggering renewed star formation. This dilution and re-enrichment process produces a younger evolutionary sequence that is separated from the older population in the age-[Fe/H] plane. In contrast, within the radial-migration prescription adopted in this work, the single-infall model produces a comparatively smooth AMR and does not reproduce the observed two-sequence structure. We therefore find that a delayed gas-accretion episode provides a straightforward explanation for the broken AMR within our modelling framework.

However, gas dilution is not the only possible interpretation of the observed AMR. \citet{Minchev_2019} proposed that the shape of the local AMR can be strongly affected by the Yule--Simpson effect. Stellar populations separated by birth radius may individually follow relatively well-defined AMRs, while their superposition in a present-day radial bin can produce a substantially flatter or even apparently non-monotonic relation. Using the high-precision LAMOST subgiant sample, \citet{Lu_2024} further showed that the apparent decline of the upper metallicity envelope over the last $\sim8$ Gyr can arise from the superposition of populations formed at different birth radii. In particular, the youngest and most metal-rich stars formed in the inner disc may not yet have had sufficient time to migrate into the solar neighbourhood. Therefore, the observed decline of the AMR upper envelope alone does not uniquely require late-time dilution of the ISM.

Our results do not rule out this interpretation. Radial migration is included in both our single- and two-infall models, although through a simplified diffusion prescription. Within this framework, radial redistribution alone is insufficient to reproduce the prominent two-sequence structure seen in the observations, whereas the inclusion of a second gas-accretion episode provides a substantially better qualitative description. Our results therefore favour recent gas accretion as an important contributor to the observed broken AMR, while more complex radial redistribution may also play a significant role. Distinguishing these processes ultimately requires a model that simultaneously treats chemical evolution and non-axisymmetric stellar dynamics in greater detail.

Our inferred gas-accretion history is broadly consistent with previous two-infall studies. \citet{Spitoni_2020}, using precise asteroseismic ages from the APOKASC sample and a Bayesian analysis, inferred a delay of approximately 4.5--5.5 Gyr between the two major accretion episodes and demonstrated that dilution associated with the second infall is important for shaping the low-$\alpha$ population. In our model, the initial rapid accretion phase is similarly followed by an extended period of secular evolution before the second accretion episode begins at a Galactic time of approximately 6 Gyr. The delayed second infall allows the older sequence to undergo substantial chemical enrichment before the younger sequence emerges.

On the other hand, \citet{Dubay_2026} showed that two-infall models can have difficulty reproducing the nearly flat AMRs inferred from APOGEE DR17 red giants. Their analysis employed stellar ages estimated using the spectroscopic neural-network method of \citet{Leung_2023}, calibrated on asteroseismic ages, as well as independent ages inferred from the [C/N]--age relation. The former typically has an age precision of $\sim22$ per cent, while the latter has an uncertainty of approximately $1.6$~Gyr, both larger than the median age uncertainty of $\sim7.5$ per cent for the subgiant sample of \citet{Xiang_2022} adopted in this work. \citet{Dubay_2026} found that strong dilution during the second accretion episode can produce metallicities that are too low for intermediate-age stars. Our model differs from theirs in the timing, duration, and metallicity of the second accretion event, as well as in the stellar sample and age estimates used for model comparison. In particular, the later onset of the second accretion episode allows more extended enrichment of the older population, while the pre-enriched infalling gas reduces the amplitude of the subsequent metallicity decrease. Nevertheless, our model also has difficulty reproducing some intermediate-age, metal-rich stars near the solar neighbourhood, indicating that gas accretion alone may not provide a complete description of the observed AMR.

\subsection{Constraints on stellar radial migration}\label{subsubsec:Constraints}
The MDFs of mono-age stellar populations provide an independent constraint on stellar radial migration. At a fixed age, stars born at different Galactocentric radii inherit different metallicities because of the radial metallicity gradient. Radial migration subsequently mixes these populations and broadens the MDF observed at a given present-day radius. Our comparison between the observed and model-predicted MDFs therefore allows us to constrain the effective strength of radial migration.

For stars with ages between 2 and 6 Gyr at $9.5<R<11.5$ kpc, we obtain a best-fitting migration strength parameter of $\sigma=2.15^{+0.100}_{-0.103}$ kpc. In terms of the migration prescription adopted in this work, this corresponds to an average radial displacement of approximately 1.92 kpc over 10 Gyr.
Our inferred migration strength is lower than several previous observational estimates. \citet{Frankel_2018} inferred an average migration distance of approximately 3.21 kpc over 10 Gyr from APOGEE red giants, while the updated analysis of \citet{Frankel_2020} obtained approximately 2.68 kpc. Using the MDFs of mono-age populations, \citet{Lian_2022} derived an average migration distance of approximately 2.74 kpc over 10 Gyr for stars born at $9<R<11$ kpc. Numerical simulations have also typically found migration distances of approximately 2.1--2.4 kpc for old stellar populations \citep{roskar_2008b,silva_2021,khoperskov_2021}.
Although the migration strength inferred in this work is lower than several previous estimates, radial migration is still required within our modelling framework to reproduce the observed broadening of the mono-age MDFs. The difference in the inferred migration strength may reflect the dependence of migration efficiency on stellar age and Galactocentric radius, as well as the different age and radial ranges probed by the stellar samples adopted in this and previous studies.

\subsection{Model limitations}\label{subsubsec:limitations}
Several limitations of the present model should be considered when interpreting these results. First, stellar radial migration is implemented using a Gaussian-like diffusion prescription based on \citet{Frankel_2018}, supplemented by a drift term that compensates for the net outward redistribution expected for a stellar disc with radially decreasing surface density. This treatment provides an effective description of stochastic radial mixing while approximately accounting for the global angular-momentum balance of the disc. However, it cannot fully describe migration driven by individual non-axisymmetric structures such as the Galactic bar, spiral arms, and their resonances. These processes can produce time-dependent, coherent, and strongly asymmetric radial redistribution. For example, \citet{zhang_2026} suggested that the slowdown of the Galactic bar can move its corotation resonance outward and preferentially transport stars towards larger Galactocentric radii.

This limitation may be related to the intermediate-age, metal-rich population that is not reproduced by our fiducial model. In the solar neighbourhood, the observations contain stars with ages of approximately 3--8 Gyr and metallicities around solar values, whereas this population is largely absent from the model prediction. \citet{zhang_2026} proposed that such stars may have formed in the inner Galaxy and subsequently migrated outward with the evolving corotation resonance of the Galactic bar. Because this type of coherent resonant migration is not explicitly represented by our diffusion prescription, the inability of the model to reproduce this population does not necessarily imply a failure of the underlying chemical evolution history.

The finite radial extent of our chemical evolution model may also affect the inferred MDFs. The current model extends only to $R=12.5$ kpc, and therefore neglects stars originating at larger Galactocentric radii. Such stars are expected to be relatively metal poor and may contribute to the low-metallicity side of the MDF in the $10.5<R<11.5$ kpc region. Extending the model to larger radii will therefore be important for obtaining more robust constraints on radial migration in the outer disc.

Finally, our treatment of the gas accretion history is highly parameterized. At a given radius, the second accretion episode is represented by a prescribed accretion rate and metallicity, whereas real gas accretion may be spatially and temporally inhomogeneous. Variations with azimuth, local mixing efficiency, and the metallicity distribution of the accreted gas could broaden the resulting stellar metallicity distribution and modify the detailed morphology of the young AMR sequence. In addition, the selection function of the \citet{Xiang_2022} sample has not been explicitly forward-modelled in the present analysis and may contribute to differences between the observed and predicted density distributions.

Future work should therefore combine a more flexible gas-accretion history with a dynamical migration model that incorporates the effects of the Galactic bar and spiral structure, extend the chemical evolution calculation to larger Galactocentric radii, and explicitly account for the observational selection function. Such improvements will allow the relative contributions of gas accretion and stellar radial redistribution to the observed AMR to be constrained more robustly.

\section{Summary} \label{sec:summary}
In this study, we use the observed distribution in age-metallicity plane, benefiting from high-precision age estimates for individual stars, to explore the chemical evolution history and stellar radial migration in the Milky Way. Observational data indicate a break in the age-[Fe/H] space of the Milky Way, which divides the stellar population into two distinct sequences. The inner disk is predominantly composed of the older sequence, while the outer disk is dominated by the younger sequence. Our results show that this break in the AMR can be reproduced by a relatively recent major gas-accretion event, although radial mixing may also contribute to shaping the observed AMR.

In our fiducial chemical evolution model, the Milky Way has experienced two phases of gas accretion that triggered two episodes of starburst. The early starburst may have been driven by merger events in the early history of the Milky Way, forming the older age-metallicity sequence. The second starburst was associated with a more recent gas accretion event, which may have been caused by the accretion of a metal-poor dwarf galaxy, built the younger sequence. The recent metal-poor gas accretion event led to a pronounced decrease in the overall metallicity of the Milky Way, giving rise to the observed break in the AMR. By adopting a modified two-infall model, we successfully reproduce both the break in the AMR and the bimodal distribution of stellar sequences in the age\text{–}[Fe/H] plane.

Furthermore, we investigate the impact of stellar radial migration on the chemical evolution of the Milky Way, and constrain its strength for stars with ages of 2–6 Gyr in the radial range of 9.5–11.5 kpc. Our model incorporates a Gaussian-like diffusion prescription for stellar radial migration based on \citet{Frankel_2018}, including a drift term for the global radial redistribution of an exponential disc, with the migration strength parameterized by \(\sigma\). We analyze the metallicity distribution function (MDF) of mono-age stellar populations in the 9.5–11.5 kpc radial range and the 2–6 Gyr age range. By comparing the MDF of mono-age populations in our model with observational data, we tested different stellar radial migration strengths ($\sigma$). We used a chi-squared test to quantify the differences and determine the best-fit result. Our findings indicate that the observational data well constrain the strength of stellar radial migration, with the best-fit migration strength being 2.15 kpc, with an uncertainty of +0.100 kpc and -0.103 kpc. This result is lower than the migration strengths reported by \citet{Frankel_2018,Frankel_2020} and \citet{Lian_2022}, suggesting that the effect of radial migration is less significant than previously thought for stars with ages of 2–6 Gyr in the Galactocentric radial range of 9.5–11.5 kpc.

\section*{Acknowledgements}
We thank the reviewer for the constructive comments and suggestions that helped improve the manuscript. We also thank Hanyuan Zhang for the insightful discussions. We acknowledge support from the National Natural Science Foundation of China (No. 12473021), the National Key R\&D Program of China (No. 2024YFA1611600), the Yunnan Province Science and Technology Department Grant (Nos. 202105AE160021 and 202005AB160002), the Yunnan Provincial Key Laboratory of Survey Science (No. 202449CE340002), and the Start-up Fund of Yunnan University (No. CY22623101).

\section*{Data Availability}

The data used in this work are taken from \citet{Xiang_2022}. The chemical evolution model framework adopted in this study is publicly available on 
\href{https://github.com/QinhaoShao/GCE_model.git}{GitHub}.  We note that the parameter settings in the publicly available version differ from those adopted in this work and therefore need to be modified accordingly.



\bibliographystyle{mnras}
\bibliography{references} 



\bsp	
\label{lastpage}
\end{document}